\documentclass[reprint]{revtex4-1}

\usepackage{graphicx}
\usepackage{amssymb}
\usepackage{amsmath}

\usepackage{ulem}

\newcommand{\Eq}[1]{Eq.~(\ref{eq:#1})}
\newcommand{\Fig}[1]{Fig.~\ref{fig:#1}}

\newcommand{\Sec}[1]{Section~\ref{sec:#1}}

\begin{document}

\title{
	Translocation through environments with time dependent mobility
      }

\author{Jack A. Cohen}
  \email[]{j.cohen@physics.ox.ac.uk}
   \affiliation{The Rudolf Peierls Centre for Theoretical Physics, University of Oxford,
 1 Keble Road, Oxford OX1 3NP, United Kingdom}
 \author{Abhishek Chaudhuri}
  \affiliation{Department of Physical Sciences, Indian Institute of Science Education and Research Mohali, Knowledge City, Sector 81, SAS Nagar, Manauli PO 140306, India}
\author{Ramin Golestanian}
 \affiliation{The Rudolf Peierls Centre for Theoretical Physics, University of Oxford,
 1 Keble Road, Oxford OX1 3NP, United Kingdom}
\date{\today}

\begin{abstract}
We consider single particle and polymer translocation where the frictional properties experienced from the environment are changing in time. This work is motivated by the interesting frequency responsive behaviour observed when a polymer is passing through a pore with an oscillating width. In order to explain this better we construct general diffusive and non-diffusive frequency response of the gain in translocation time for a single particle in changing environments and look at some specific variations. For two state confinement, where the particle either has constant drift velocity or is stationary, we find exact expressions for both the diffusive and non-diffusive gain. We then apply this approach to polymer translocation under constant forcing through a pore with a sinusoidally varying width. We find good agreement for small polymers at low frequency oscillation with deviations occurring at longer lengths and higher frequencies. Unlike periodic forcing of a single particle at constant mobility, constant forcing with time dependent mobility is amenable to exact solution through manipulation of the Fokker-Planck equation.
\end{abstract}

%\pacs{}

\maketitle

\begin{figure*}[htbp]
\centering
\includegraphics[bb=0 0  481 155]{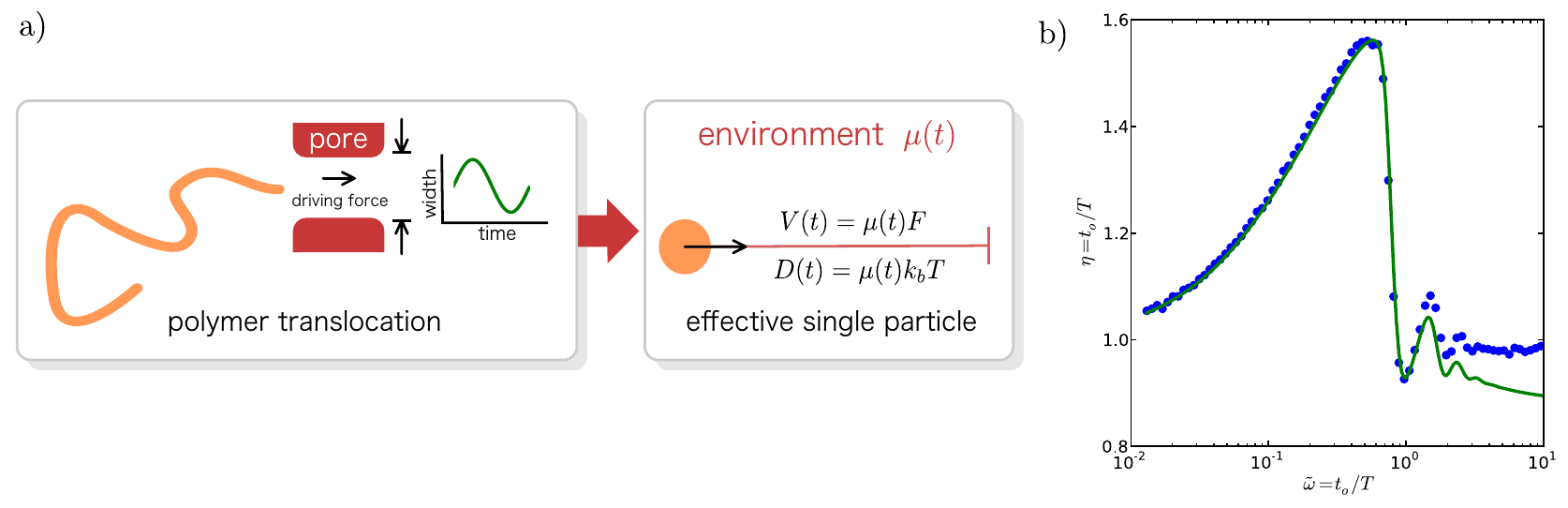} 
\caption{ a) The time taken for a polymer to move through a pore with a width that is changing harmonically in time is dependent on the frequency of the oscillation. In order to better understand this process we present an effective description of the translocation dynamics using a Fokker-Plank description with a time dependent mobility. We develop the idea of single particle time dependent mobilities in a general setting to allow for greater applicability to other soft matter systems. b) The scaled frequency response for the gain in polymer translocation time from simulations (dots) is reproduced well at low frequencies
by the effective single particle description (solid line) . However, at higher frequencies the response is only matched qualitatively. This is shown for simulation parameters $N=128$, $\epsilon = 1$ and $F = 1$ with a width that varies between $2.0\sigma$ and $2.5\sigma$ sinusoidally where $\sigma$ is the particle diameter. The reference time $t_o$ is chosen to to be the translocation time at $W = 2.25\sigma$.
 \label{fig:sch}}
\end{figure*}

\section{Introduction}
The study of confinement effects in soft matter systems has gained considerable
importance in recent years \cite{Binder2008}. The length scales of these systems 
such as colloids and polymers are in the mesoscale, making experimental study involving single particle tracking feasible and hence the possibility of understanding the 
microscopic effects of confinement better \cite{Poon2002,Kerle1996,Kerle1999,Blaaderen1997,Aarts2003,Aarts2004}.  For example, there have been efforts to understand the changing phase behaviors of confined soft fluids by altering the interaction of the particles of the fluid with the confining walls \cite{Lowen2001,Halperin1991,Grest1995,Granick1999,Advincula2004}. { It is interesting to ask how such fluids would behave if the confinement is changing in a time dependent fashion.}

Confinement  induced by external fields has been realised and studied experimentally by techniques such as optical tweezers and crossed laser beams. {In bulk colloidal suspensions the use
of crossed laser beams forces a spatially oscillating external potential and leads to the 
phenomenon of laser-induced freezing, and has been extensively studied both in experiments
and using computer simulations \cite{Chowdhury1985,Loudiyi1992,Wei1998,Bechinger2000}.} {Indeed,} simulations have \sout{also} revealed interesting behaviours when the external potential and hence the confinement is time dependent. 
However, the effect of time dependent spatial confinement on particles in a flow, such as colloids or polymer translocation through channels, has not yet received much attention. Since a wide range of complex synthetic and biological channels are becoming available for experimentation it is now relevant to understand the impact  of time dependent changes in these environments on translocation;  this is the focus of the present study. Such time dependent changes could lead to frequency dependent selectivity of different targets, which could be important in micro particle sorting or identification devices.

This work is motivated by the interesting behaviour observed in the frequency response of the translocation time of a polymer moving through a pore with a width oscillating in time \cite{Cohen2011}, see Figure \ref{fig:sch}. We find this behaviour can be reproduced qualitatively by considering a mobility that varies in time. We develop the idea of time dependent mobilities in general to allow for wider applicability to other soft matter systems with time varying confinement returning to the specific case of polymer translocation at the end of the article. 
% We consider the effect of time dependent mobilities in general to allow a wider applicability to other soft matter systems with time dependent confinement. }
%consider mobilities that vary in time in order to understand this behaviour.
%{ construct a Fokker-Planck equation to obtain the first-passage time distribution with a time dependent mobility capturing the changes in the environment due to the oscillating cross-section of the pore. Many of these features can be explained by simple models excluding noise. We develop these ideas for general systems by introducing these simple deterministic models, and then include the effect of noise.  }

Unlike Brownian motion in homogeneous media, the diffusion coefficient, $D$, of a particle in the presence of barriers such as membranes and confining walls is no longer constant in time even though the mobility of the particle remains unchanged \cite{Haus1987,Bouchaud1990,Aronovitz1984,Novikov2011}. This is due to the mean squared displacement departing from a simple linear dependence on time at times much greater than the molecular time scale. Alternatively, time dependent changes of the environment, such as contraction of the surrounding geometry, could immediately affect the mobility of the particle, also resulting in a time dependent diffusion coefficient. This could arise from changing bulk or surface frictional properties of the environment  prompted by changes in external chemical, electromagnetic or mechanical stress fields.

%CHANGE
In this study we restrict ourselves to situations where the friction experienced by the particle changes in a time dependent fashion {\it homogeneously} across the entire system. Within the linear response regime, a particle acted on by a constant external force, $F$, will experience the velocity $V(t) = \mu(t) F$, we assume the corresponding diffusion coefficient $D(t) = \mu(t) k_bT$.  In this restriction, both drift and diffusion have the same time dependence with a constant ratio between them. This is a convenient property that allows analytical solution of the Fokker-Planck equation \cite{Molini2011}.

In this paper we investigate, both analytically and using computer simulations, the dynamics of translocation of a single particle through an environment that is oscillating in a time dependent manner. We look at both diffusive and non-diffusive translocations of the particle. In  non-diffusive translocation, we look at a basic case of of two state confinement in the absence of noise:  the particle is either moving with a steady velocity or is stationary. We then generalise this for any time periodic confinement and discuss specific cases. In diffusive translocation, we solve the relevant Fokker-Planck equation with time dependent mobility and construct the general first passage time distribution for any time dependent confinement. We finally compare the results of single particle translocation to the more  technologically and biologically relevant problem of polymer translocation in the presence of such time varying confinement. 

The rest of the paper is organised as follows, in \Sec{twostate} we construct the frequency response of the gain in translocation time for a deterministic particle in two state confinement, that either moves with a constant velocity or is stationary. We allow for asymmetric `on' and `off' intervals. In \Sec{genperiodic} we extend this analysis to any periodic variations in the mobility and present the general form for the deterministic gain. We apply this to two specific cases of sawtooth and sinusoidal variations and compare the responses. In \Sec{diffusivegain} we introduce diffusion and derive the first passage time distribution for an arbitrary time dependent mobility. We then construct an exact solution for the diffusive counterpart of the two state confinement of \Sec{twostate}. We compare the diffusive two state confinement to sinusoidal variations in the mobility for different relative strengths of diffusion. In \Sec{numosc} we numerically integrate the stochastic equation of motion for time dependent mobilities at constant driving force and match the results to the analytical form. \Sec{polymer} then applies this single particle description to a numerical model of polymer translocation through a pore with time dependent width.

\begin{figure}[htbp]
\centering
\includegraphics[bb=0 0  252 252]{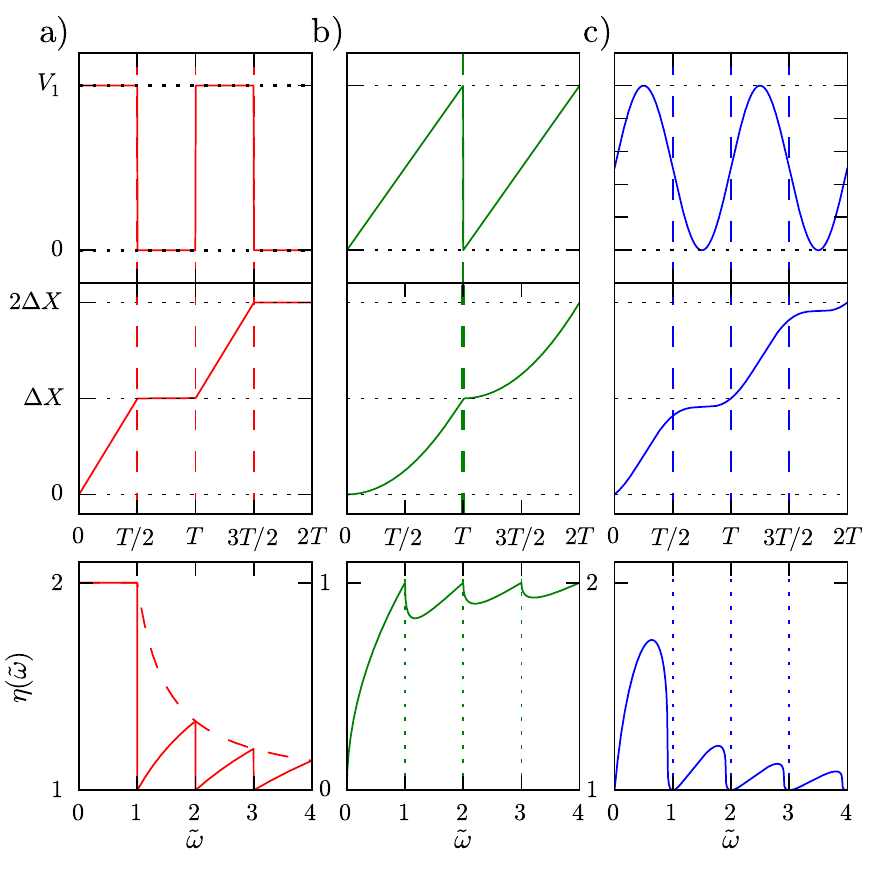} 
\caption{ Time dependent velocity (top) and position (vertical middle) as a function of the time period of oscillation for (a) two state (b) sawtooth and (c) sinusoidal mobilities. (Bottom) Frequency response of the gain. The gain is shown for four dimensionless frequencies}
 \label{fig:2state}
\end{figure}

\section{Non diffusive translocation : Two state confinement}
\label{sec:twostate}

We first consider a situation where a particle under the influence of a constant driving force in one dimension experiences a two state confinement: a state where it is able to move with constant velocity ($V_{{\rm on}}$) and the other where it is stuck  with zero velocity ($V_{\rm off} = 0$). Therefore, the mobility switches between a constant value ($\mu_{{\rm on}}$) and zero. This could occur, for example, if a charged colloid moves  due to an external electric field through a long channel that becomes strongly oppositely charged, or the channel is mechanically compressed in a transverse direction, such that the colloid  is temporarily fixed to the surface without the possibility of slipping. Since the durations of the on ($t_{\rm on}$) and off ($t_{\rm off}$) states of confinement need not be symmetric, we define the  asymmetry parameter $a = t_{\rm off}/t_{\rm on}$ over the period $T (= t_{\rm on} + t_{\rm off})$. To analyse the  dynamics, we look at the efficiency of the translocation process by comparing the time taken for the particle to traverse a distance $L$ in the presence of this oscillatory two-state confinement, $t_{\rm osc}(\tilde{\omega})$, to one where the particle moves unhindered with a velocity equal to the average over an oscillation period with symmetric off and on durations, $t_0=\langle t\rangle_{T,a=1}$. This efficiency is defined as the gain in translocation rate
\begin{eqnarray}
\eta(\tilde{\omega}) &=& \frac{t_0}{t_{\rm osc}(\tilde{\omega})}
\end{eqnarray}
where $T$ is the time period of oscillation and $\tilde{\omega} = t_0/T$ is a dimensionless frequency.  The average velocity of the particle in peroid $T$ with $a=1$ is $\langle V\rangle_{T,a=1} = V_{\rm on}/2$, and so $t_0 = 2L/V_{\rm on}$. 

The time taken by the particle to translocate a distance $L$ is 
\begin{eqnarray}
t_{\rm osc} &=& N_{\rm off} t_{\rm off} + N_{\rm on} t_{\rm on} \label{eq:tosc2state}
\end{eqnarray} 
where $ N_{\rm off}$ and $N_{\rm on}$ are the number of off and on cycles respectively. Note that since the particle only moves in the on state, $N_{\rm on}$ can be fractional while $N_{\rm off}$ is an integer. Also, $L = V_{\rm on} N_{\rm on} t_{\rm on}$ which implies, $t_0 = 2N_{\rm on}t_{\rm on}$. Using these relationships we can express the gain as a function of the number of `off' to `on' cycles as, 
\begin{eqnarray}
\eta &=& 2 \left[1 + a \frac{N_{\rm off}}{N_{\rm on}}\right]^{-1}
\end{eqnarray}
We can rewrite $\eta$ in a continuum form by expressing the off cycles as $N_{\rm off} = \mathcal{S}(N_{\rm on})$, where $\mathcal{S}(x)$ is the unit staircase function and returns the integer value of $x$. More details of the staircase function are listed in the Appendix. Since $N_{\rm on} = \tilde{\omega}$, we have
\begin{eqnarray}
\eta(\tilde{\omega}) &=& 2 \left[1 + \frac{2a}{1+a} \frac{\mathcal{S}((1+a)\tilde{\omega}/2)}{\tilde{\omega}}\right]^{-1} \label{eq:2state}
\end{eqnarray}
For any value of $a$ the gain is at a maximum when $N_{\rm off} = N_{\rm on} - 1$ and at a minimum when $N_{\rm off} = N_{\rm on}$, as shown in Fig ~\ref{fig:2state}a. For symmetric cycles the gain takes the simple form $\eta(\tilde{\omega}) = 2 \left[1 +  \frac{\mathcal{S}(\tilde{\omega})}{\tilde{\omega}}\right]^{-1}$ with maximum $\eta_{\rm max}(\tilde{\omega})=2\tilde{\omega}/(2\tilde{\omega}-1)$ and minimum $\eta_{\rm min} = 1$. We also note that  $\eta(\tilde{\omega} \rightarrow \infty) \rightarrow 2/(1+a)$. 

\section{Non diffusive translocation : General periodic confinement}
\label{sec:genperiodic}

It is possible to generalise the process of obtaining the frequency response of the gain for any periodic time dependent mobility, $\mu (t)$, in the presence of the constant driving force, $F$, to find the following expression 
\begin{eqnarray}
\eta(\tilde{\omega}) &=& \left[ \frac{\mathcal{S}\left(\tilde{\omega}\right)}{\tilde{\omega}} + \Delta \tilde{t}_{\rm osc}\left(1 - \frac{\mathcal{S}(\tilde{\omega})}{\tilde{\omega}}\right)\right]^{-1} \label{eq:ndpc}
\end{eqnarray}
where $\Delta \tilde{t}_{\rm osc}(\Delta L)$ is the time needed to translate the remaining distance, $\Delta L$, occurring after the complete cycles. Further information relating to the construction of this expression can be found in Appendix \ref{sec:appb}

\subsection{Sawtooth mobility}
For the case of a simple sawtooth mobility $\mu(t) = \mu_0 t /T$, the average velocity being $\langle V \rangle_T = \mu_0F/2$, where $\mu_0$ is a constant amplitude. Therefore,  $\tilde{X}(\tilde{t}) = \tilde{\omega}{\tilde{t}}^2$ and $\Delta \tilde{t}_{\rm osc}(\tilde{X}) = (\tilde{X}/\tilde{\omega})^{1/2}$ with gain
\begin{eqnarray}
\eta(\tilde{\omega}) &=& \frac{\tilde{\omega}}{\mathcal{S}\left(\tilde{\omega}\right) + \sqrt{\tilde{\omega} - \mathcal{S}(\tilde{\omega})}}
\end{eqnarray}
The minimums occur at $\tilde{\omega} = 2(n+n^2-\sqrt{n^3 + n^4})$ for period $n>0 \in \mathbb{Z}$ and maximums $\eta_{\rm max} = 1$ for $\tilde{\omega} = n$ as shown in Fig. ~\ref{fig:2state}b.

\subsection{Sinusoidal mobility}

For an oscillating confinement giving rise to a sinusoidal mobility as $\mu(t) = \mu_0(1+\sin(2\pi t/T))/2$, the average velocity over a period is $\langle V\rangle_T = \mu_0F/2$. In this case, $\tilde{X}(\tilde{t}) = {\tilde t} + (1-\cos(2\pi\tilde{\omega}{\tilde t}))/(2\pi\omega)$ with $\tilde{X}(0)=0$. Thus to solve for $\Delta \tilde{t}_{\rm osc}$, we need to solve a transcendental equation, which we do numerically. The gain plots are shown in Fig. ~\ref{fig:2state}c. 

\section{Diffusive Translocation : General Treatment}
\label{sec:diffusivegain}
We now incorporate the effect of diffusion in the translocation dynamics of a particle experiencing time dependent confinement. We consider a particle in 1D initially located at some distance $L$ from an absorbing barrier at the origin. An external force drives the particle towards the barrier with a drift velocity $V= \mu(t)F$. The particle traverses the distance $L$ before absorption and during this process we assume it experiences a time dependent confinement that manifests in the diffusion coefficient instantly as $D = \mu(t)k_bT$.  

We first study the case of a  general confinement and determine the  first passage time distribution, from which we can obtain the translocation time as first mean passage time. Then we study specific cases and look at the frequency response of the gain $\eta(\tilde{\omega}$) in the translocation process and finally compare with our previous study to investigate the effect of diffusion.

To do so, we write down the Fokker-Planck equation for this process as
\begin{eqnarray}
\frac{\partial p}{\partial t} &=& \mu(t)\left( k_bT \frac{\partial^2 p}{\partial x^2} + F\frac{\partial p}{\partial x} \right)
\end{eqnarray}
Choosing $L$, $L/V$ and $\mu_0$ for the units of length, time and mobility we can rewrite this equation in dimensionless parameter form
\begin{eqnarray}
\frac{\partial p}{\partial \tilde{t}} &=& \tilde{\mu}(t)\left( {\tilde{f}}^{-1} \frac{\partial^2 p}{\partial \tilde{x}^2} + \frac{\partial p}{\partial \tilde{x}} \right).
\end{eqnarray}
$\tilde{f} = LV/D = FL/k_b T$ is a dimensionless force measuring the work done against thermal energy, with the effect of diffusion reducing with increasing $\tilde{f}$, and $\mu_0 = \mu_{\tilde{\omega}=0}$ is some reference mobility chosen to be the mobility when the confinement is static. The positivity of the drift term is a result of the force being directed towards the origin from an initial position $L$. This equation can be transformed to the time independent mobility form by a change of variable $p(\tilde{x},\tau(\tilde{t}))$ \cite{Molini2011} where
\begin{eqnarray}
\tau(\tilde{t}) = \int_0^{\tilde{t}} \tilde{\mu}(\tilde{t}^\prime)d\tilde{t}^\prime 
\end{eqnarray}
 The transformation yields
\begin{eqnarray}
\frac{\partial p}{\partial \tau} &=& {\tilde{f}}^{-1} \frac{\partial^2 p}{\partial \tilde{x}^2} + \frac{\partial p}{\partial \tilde{x}} 
\label{eq:taut}
\end{eqnarray}
This equation can be once more transformed into a drift free equation by the substitution $p(\tilde{x},\tau) = q(\tilde{x},\tau)w(\tilde{x})s(\tau)$ where $w = \exp\left(-{{\tilde{f}}\,\tilde{x}}/{2}\right)$ and $s = \exp\left(-{{\tilde{f}}\,\tau}/{4}\right)$ \cite{Schuss2010} giving
\begin{eqnarray}
\frac{\partial q}{\partial \tau} &=& {\tilde{f}}^{-1} \frac{\partial^2 q}{\partial \tilde{w}^2} 
\end{eqnarray}
The initial condition satisfying the absorbing boundary condition at the origin is now $q(x,0) = \exp({ \tilde{f}\,L}/2)(\delta(x-L)-\delta(x+L))$. Solving for $q$, transforming to $p$, making the change of variable, $\tau(\tilde{t})$, and a little rearrangement leads to 
\begin{eqnarray}
\tilde{p}(\tilde{x},\tilde{t}|1) &=& \frac{e^{-\frac{(\tilde{x}-1+\tau(\tilde{t}))^2}{4{\tilde{f}}^{-1}\tau(\tilde{t})}} - e^{\tilde{f}}e^{-\frac{(\tilde{x}+1+\tau(\tilde{t}))^2}{4{\tilde{f}}^{-1}\tau(\tilde{t})}}}{{\sqrt{4\pi {\tilde{f}}^{-1}\tau(\tilde{t})}}}
\end{eqnarray}
Defining the survival probability, $\tilde{P}(\tilde{t})=\int_0^\infty p(\tilde{x},\tilde{t})d\tilde{x}$, we  find
\begin{eqnarray}
\tilde{P}(\tilde{t}) &=& \frac{1}{2}\left[ {\rm erfc}\left( \frac{\tau(\tilde{t})-1}{\sqrt{4{\tilde{f}^{-1}\tau(\tilde{t})}}}\right)-e^{\tilde{f}} {\rm erfc}\left( \frac{\tau(\tilde{t})+1}{\sqrt{4{\tilde{f}^{-1}\tau(\tilde{t})}}}\right)\right] \nonumber
\end{eqnarray}
 with the first passage time distribution $\tilde{\rho}=-\partial_{\tilde{t}}\tilde{P}(\tilde{t})$, as
\begin{eqnarray}
\tilde{\rho}(\tilde{t}) &=& \frac{\tilde{\mu}(\tilde{t})}{\sqrt{4\pi{\tilde{f}}^{-1}\tilde{\tau}(\tilde{t})^3}} \exp\left({- \frac{(1-\tau(\tilde{t}))^2}{4{\tilde{f}}^{-1}\tau(\tilde{t})}}\right)
\label{eq:rhoosc}
\end{eqnarray}
The translocation time is obtained as the mean first passage time, $\langle \tilde{t} \rangle = \int_0^{\infty}\tilde{t} \tilde{\rho}(\tilde{t})d\tilde{t}$. For the special case of time independent mobility we have  $\tilde{\mu}=1$ and $\tilde{\tau}(\tilde{t})=\tilde{t}$ and in rescaled units the mean first passage time takes the convenient form $\langle \tilde{t} \rangle = 1$.

\begin{figure*}[htbp]
\centering
\includegraphics[bb=0 0 494 156]{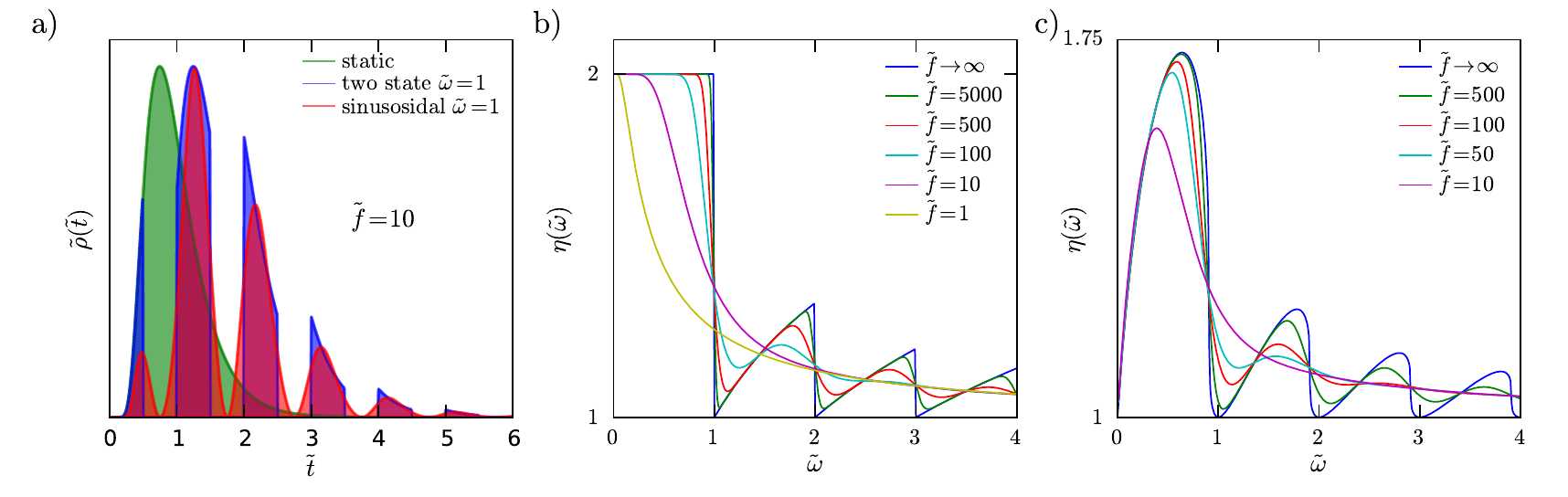} 
\caption{ a) Comparison of first passage time distributions for static non oscillating confinement, two state confinement and sinusoidal confinement with dimensionless frequency $\tilde{\omega}=1$  at ${\tilde{f}}=10$.  The two state confinement distribution is constructed by inserting regions of zero probability in the static confinement distribution and shifting the distribution appropriately by the off time. The minimums of the sinusoidal confinement occur at $\tilde{t} = n + 0.75$ where $n$ is a positive integer, corresponding to the minimum in the cycle. b) Frequency response of the gain for two state confinement. As ${\tilde{f}}\rightarrow\infty$ the frequency response matches that of the deterministic case as expected. At low ${\tilde{f}}$ the peaks become smoothed and are eventually washed out in the highly diffusive limit. c) The frequency response of the gain for sinusoidal confinement behaves similarly to the two state confinement. Both cases show an enhancement in translocation over the average velocity of each cycle.\label{fig:diff}}
\end{figure*}

\subsection{Two state confinement}

For the two state confinement case the mobility either take values $\mu_{\rm off} = 0$ or $\mu_{\rm on} \ne 0$. During periods of zero mobility the particle will be stationary without any diffusion and there will be regions of zero probability in the first passage time distribution. The mobility takes on the form of a periodic square wave, as in \Fig{2state}a. Integration of the square wave mobility gives the effective time $\tau$ (Eq. ~\ref{eq:taut}), which experiences plateaus corresponding to stops during the off cycle. We may rewrite the first passage distribution by including these stoppage times with an appropriate shift in time
\begin{eqnarray}
\langle t\rangle_{\rm osc} &=& \sum_{n=0}^{\infty} \int_{n t_{\rm on}} ^{(n+1)t_{\rm on}} (t + n t_{\rm off}) \rho(t) dt\\
&=& \langle t\rangle_{\rm static} + t_{\rm off}\sum_{n=0}^{\infty} \int_{n t_{\rm on}} ^{(n+1)t_{\rm on}} n \rho(t) dt
\end{eqnarray}
where $\langle t\rangle_{\rm static} = L/V_{\rm on} = \langle t\rangle_T/2$. The last integral may be written as a convolution with a staircase function $\mathcal{S}(x)$
\begin{eqnarray}
\langle t_{\rm osc}\rangle &=& \frac{\langle t \rangle_T}{2}  + t_{\rm off}\int_0^\infty \mathcal{S}\left(\frac{t}{t_{\rm on}}\right) \rho(t)dt\label{eq:tosc2stateS}
\end{eqnarray}
Comparison to \Eq{tosc2state}, suggests that the integral can be interpreted as the {\it average} number of off cycles before translocation
\begin{eqnarray}
\langle N_{\rm off} \rangle &=& \int_0^\infty \mathcal{S}\left(\frac{t}{t_{\rm on}}\right) \rho(t)dt
\end{eqnarray}
Using integration by parts, noting that $\mathcal{S}(0)=0$, $\mathcal{S}^\prime(x) = \sum_{n=1}^\infty \delta(x-n)$ and the survival probability $P(t)  \rightarrow 0 $ as $t\rightarrow \infty$, we  find
\begin{eqnarray}
\langle N_{\rm off}\rangle&=& \sum_{n=1}^{\infty} P\left(n t_{\rm on}\right)
\end{eqnarray}
Now, as before we can let $a=t_{\rm off}/t_{\rm on}$ and $t_{\rm on} = \langle t \rangle_T/2N_{\rm on}$ and rewrite \Eq{tosc2stateS} in terms of the gain
\begin{eqnarray}
\eta &=& 2 \left[1 + a \frac{\langle N_0 \rangle}{N_1}\right]^{-1}
\end{eqnarray}
Using the dimensionless form of the survival probability, $\tilde{P}(n\tilde{t}_{\rm on})$, we can express the gain as a function of $\tilde{\omega}$,
\begin{eqnarray}
\eta(\tilde{\omega}) &=& 2 \left[1 +  \frac{2a}{(1+a)\tilde{\omega}}\sum_{n=1}^{\infty}\tilde{P}\left(\frac{2n}{(1+a)\tilde{\omega}}\right)\right]^{-1}
\end{eqnarray}
Therefore, the gain is a function of $a$, $\tilde{\omega}$ and ${\tilde{f}}$. Fig. \ref{fig:diff} shows clearly that there is substantial gain in the translocation process for finite oscillation frequencies when compared to static confinement with a mobility that is an average over a period of oscillation. The gain oscillates with the oscillation of the confinement  decaying to $1$ as $\omega\rightarrow\infty$. Thus, translocation in a time dependent confinement in the presence of a constant driving force, can enhance the translocation process at low frequencies, as compared to the cycle average, despite stationary periods. With higher $\tilde{f}$, the plot moves to the non-diffusive regime as expected. Note that for smaller $\tilde{f}$ the oscillations smoothen out.  Diffusion cannot occur in the off periods due to zero mobility, however the diffusive behaviour in the on periods can cause translocations to complete at times around the deterministic time, smoothing the sharp drop that occurred at integer $\tilde{\omega}$. As the diffusion becomes dominant the first passage time distribution becomes broader, increasing the mean first passage time, which in turn decreases the gain. For a sinusoidal confinement, the situation is similar with  smoother oscillations than the two state confinement.

\begin{figure}[h!]
\centering
\includegraphics[bb=0 0 252 433]{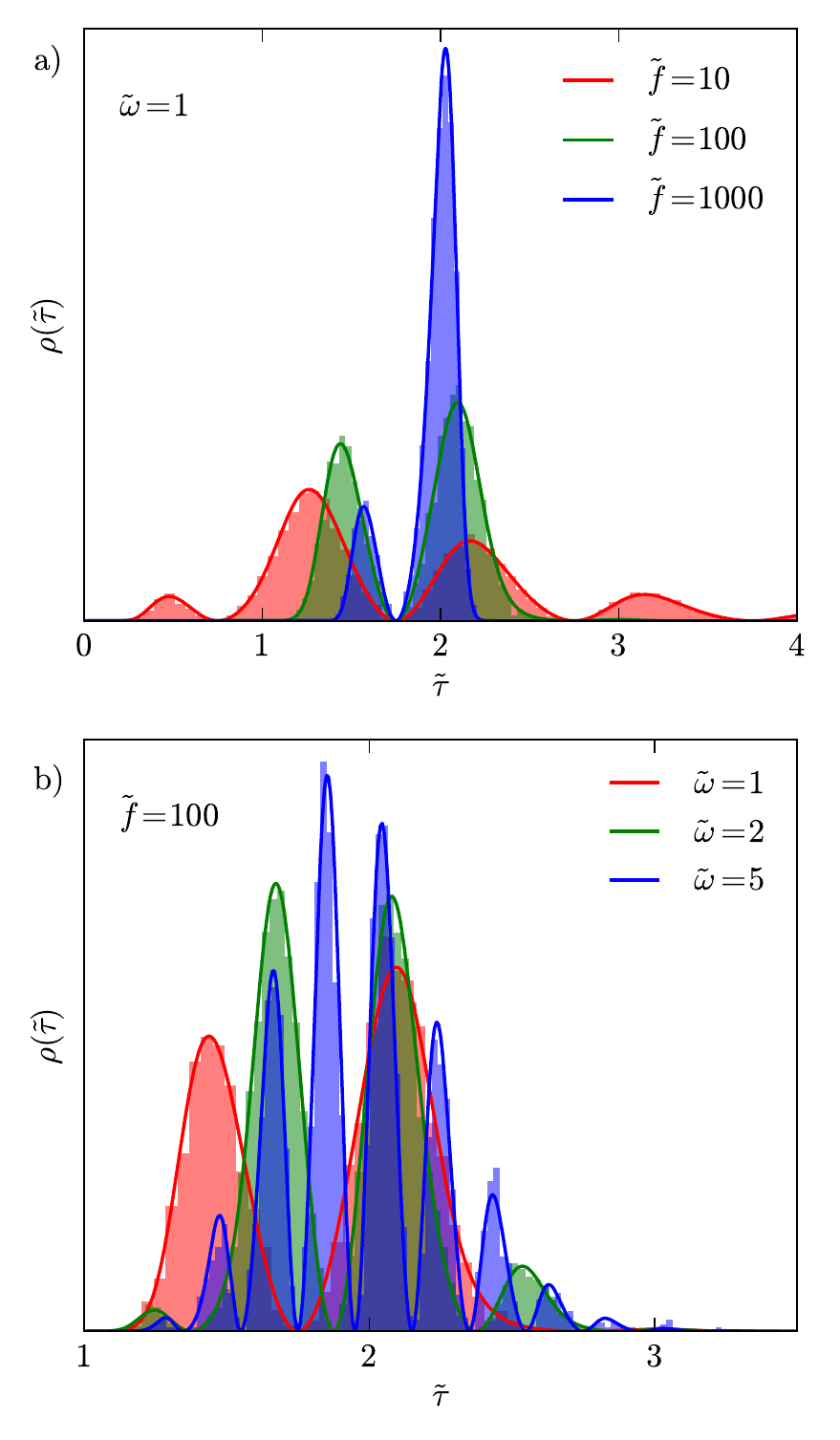} 
\caption{ Comparison between the analytical result for first passage time with time dependent mobility \Eq{rhoosc} with the results from numerical integration of the corresponding differential equation. The analytical form is plotted by the bold solid line while the histogram of numerically obtained first passage times are shown by the filled areas. The mobility varies sinusoidally between zero and its maximum. The effect of varying the dimensionless force with constant number of oscillations per non-oscillating translocation time, $\tilde{\omega}=1$, is shown in the top panel and the effect of increasing $\tilde{\omega}$ at constant ${\tilde{f}}=100$ is shown in the bottom panel.
 \label{fig:single}}
\end{figure}

\section{Numerical solution of oscillating mobilities}
\label{sec:numosc}

We now compare the analytical result for the first passage time distribution \Eq{rhoosc}, with numerical simulations. To do this, we numerically integrate the following stochastic equation of motion for a single particle,
\begin{eqnarray}
\nonumber
dX &=& V(t) dt + \sqrt{2D(t)dt}R(t)\\
	&=& \mu(t)F dt + \sqrt{2\mu(t) k_bT dt}R(t)
\end{eqnarray}
where $R(t)$ is a delta correlated white noise random number with unit variance. In dimensionless units,  this becomes
\begin{eqnarray}
d\tilde{X} &=& \tilde{\mu}(\tilde{t})d\tilde{t} + \sqrt{2\tilde{\mu}(\tilde{t}){\tilde{f}}^{-1}d\tilde{t}}R(t)
\end{eqnarray}
For sinusoidal confinement, we choose $\tilde{\mu}(\tilde{t})=(1+\sin(2\pi\tilde{\omega}\tilde{t}))/2$ as in our analytics. Using this scheme, we evaluate the time taken by a particle to traverse a distance $L$ with the given mobility. Using a number of simulation runs, we construct the translocation time distribution, which we compare with the first passage time distribution given by \Eq{rhoosc}. The comparisons are made for a set of values of ${\tilde{f}}$ and $\tilde{\omega}$. In \Fig{single} we show the effect of varying ${\tilde{f}}$ at constant $\tilde{\omega}$ and vice-versa. The results are in excellent agreement confirming the analytical result. Increasing the dimensionless force results in narrowing the distribution while keeping the number of minimums fixed. This is because at higher dimensionless force diffusion is limited. Increasing $\tilde{\omega}$ at constant ${\tilde{f}}$ splits the distribution into an increasing number of peaks.

\section{Polymer translocation through an oscillating nanopore}
\label{sec:polymer}

\begin{figure*}[htbp]
\centering
\includegraphics[bb=0 0 494 252]{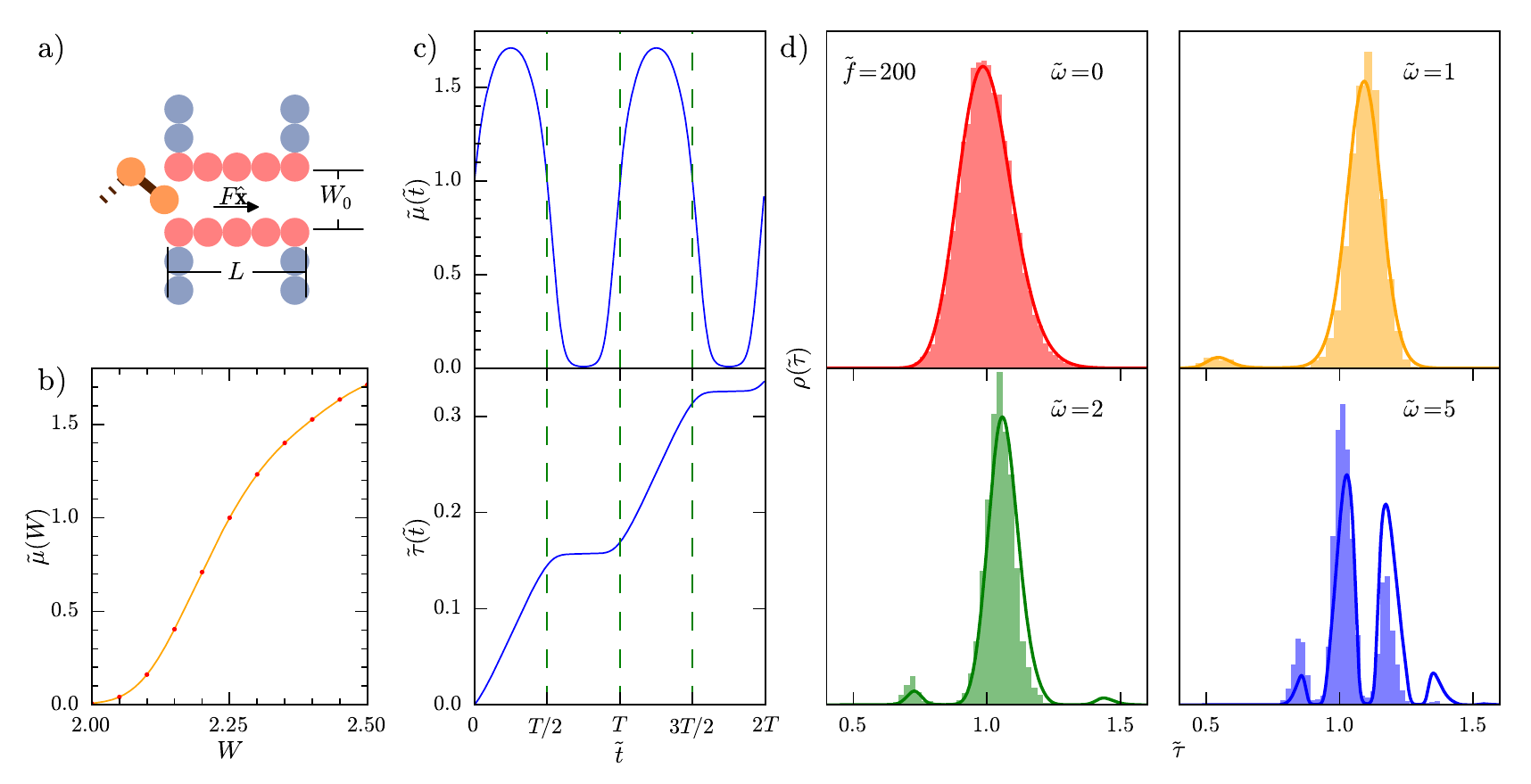} 
\caption{ a) Schematic diagram of the simulation geometry with the polymer entering from the left. b) The mobility as a function of width for simulation parameters $\epsilon = 1$, $F=1$ and $N=128$. c) The mobility and its integral, $\tilde{\tau}$, as a function of time for sinusoidal variation of the width. d) First passage time distributions for different values of the scaled frequency, $\tilde{\omega}$. As the scaled frequency increases the quasi-static approximation breaks down. The polymer has a higher probability of translocating faster as compared to the distribution  obtained from the single particle picture.
 \label{fig:polymer}}
\end{figure*}

Having established the results for translocation of a single particle in the presence of an external drive in a time dependent confining geometry, we now ask how the results compare to the {\it collective} motion of particles in a similar scenario. This question is especially relevant in the context of polymer translocation. The translocation of a polymer consisting of several monomers through a nanopore has been the subject of intense study for the last decade \cite{MuthukumarBook}. Apart from its obvious relevance in several biological processes such as RNA transport and virus ejection, it has   technological implications in problems such as DNA sequencing and drug delivery. There have been  many experimental and theoretical works in this field \cite{Salman2001,Meller2003,Meller2001,Kasianowicz1996,Akeson1999,Sauer2003,Storm2005,Sung1996,Muthukumar1999,Lubensky1999,Metzler2003,Kantor2001,Kantor2004, Milchev2004,Kantor2011,Luo2007,Sakaue2007,Sakaue2011,Gerland2004,Slonkina2003,Gopinathan2007,Dorfman2010,Frank2010,Frank2000,Sakaue2006,Sakaue2005,Asfaw2010,Lubensky1999,Cohen2012} trying to understand the translocation process and how the pore geometry and the affect of pore-polymer interactions. Although recent simulation work has shown that oscillations of the pore significantly affect the translocation process, and is especially influential in the high frequency regime, the origin of the observed phenomenon is not clear. Here we compare the earlier simulation results with our analytics for a single particle to clarify the scenario. 

As in ref \cite{Cohen2011}, we consider a bead-spring polymer in two dimensions consisting of $N$ monomers translocating through a pore whose width changes in time during the translocation process. The time dependence of the width is given by $W(t) = W_0 + W_A \sin(\omega t)$, where $W_0$ is the average width and $W_A$ is the amplitude of oscillation. The polymer inside the pore is driven by a constant force, $F$,  which drags the polymer from the {\it cis} to the {\it trans} side. The pore in this picture is made up of monomers of diameter $\sigma$. The pore-polymer interactions are modelled by an attractive Lennard-Jones potential $U_\mathrm{pm}^\mathrm{LJ}(r) = 4\epsilon_\mathrm{pm} \left[ \left(\frac{\sigma}{r} \right)^{12} - \left(\frac{\sigma}{r}\right)^{6} \right]$ for $r \le 2.5\sigma$ and $0$ for $r > 2.5\sigma$. The dynamics of the polymer is studied by integrating the equation of motion of the polymer beads
\begin{equation}
m{\bf \ddot{r}_i} = - {\mbox{\boldmath$\nabla$}} U_i + \mathbf{F}_\mathrm{e} - \xi {\bf v}_i + {\mbox{\boldmath$\eta$}}_{i}
\end{equation}
where $m$ is the mass of the monomer, $U_i$ is the total potential acting on a bead which includes the excluded volume interaction between the monomer beads of the polymer and with the wall, the FENE potential acting between consecutive monomer beads and the pore-polymer interaction, $\xi$ is the friction coefficient, ${\bf v}_i$ is the monomer velocity, and ${\mbox{\boldmath$\eta$}}_i$ is the random force satisfying the fluctuation dissipation theorem.

We look at the time taken by a polymer starting with all beads in {\it cis} side to cross the the {\it trans} side, $\tau$. To determine the time dependent mobility of the polymer in the presence of the oscillating pore, we determine the translocation times for the polymer at fixed pore widths and construct the translocation rate, $k = N/\tau_{\rm stat}$, as a function of the width, $W$. Defining the mobility rate of the polymer as $\mu = k/F$, we  have the mobility as a function of width as shown in \Fig{polymer}(b). A  cubic interpolation of this variation gives $\tilde{\mu}(W)$, and since we know the variation of the pore width as a function of time, we get an expression for the time dependent mobility of the polymer through the oscillating pore. We then substitute this expression for $\tilde{\mu}(W(\tilde{t})) \equiv \tilde{\mu}(t)$, in \Eq{rhoosc} to  obtain the first passage time distribution of the polymer through the pore. We compare the results with the simulation results for translocation time distributions of the polymer through the pore as it oscillates periodically. Note that we have used the single particle result for the first passage time distribution to predict results for the polymer problem. Although this is a major assumption, we believe that qualitatively the results should be similar. 

Indeed, we find that for a long polymer translocating through a weakly attractive {\it static} pore, $\tilde{\omega}=0$, the distribution is matched well by ${\tilde{f}}=200$. Since we do not have an estimate of $\tilde{f}$ for the polymer problem, we use this value for the dimensionless force to measure the mobility as a function of pore width in \Eq{rhoosc}. The comparison of the first passage time distribution as obtained from the simulation with the single particle description is shown in \Fig{polymer}d) and the gain as calculated from the mean first passage time is shown in \Fig{sch}b). It can be seen that the description matches well for small values of $\tilde{\omega}$ but as the value increases deviations occur, with the polymer favouring translocation at shorter times than predicted from the single particle picture.

\section{Conclusion and discussion}

We have investigated the effect of time dependent mobility on single particle translocation and compared these results to those obtained from polymer translocation simulations through attractive pores. For deterministic translocation of a single particle we constructed the frequency response of the gain for the two state system. This exhibited clear features as a result of the `off' period, with a `ringing' gain ultimately decaying to unity as the dimensionless frequency tends to infinity. Developing from this, we formulated the frequency response for any time periodic changes in the mobility. We applied this formalism to a sawtooth ramping mobility and sinusoidal oscillations. 
We moved on to include diffusion in the translocation process and found an exact expression for the gain in the two state problem and numerical solutions to the sinusoidally oscillating mobilites. These results were compared against the deterministic case by taking the limit ${\tilde{f}}\rightarrow\infty$.

The simple single particle picture was then applied to the problem of polymer translocation through a pore with a periodically time varying width. A time dependent mobility was constructed by numerically measuring the average mobility at different widths and then modulating the mobility in time with the prescribed sinusoidal variation of the width.  Better agreement is found for a polymer of length $N=32$ and $N=64$ translocating through a weakly attractive pore (data not shown). For $N=128$, see \Fig{polymer}d, it can be seen that the single particle description is overestimating the translocation time at $\tilde{\omega}=5$, suggesting that the entropic nature of the polymer is in fact speeding up the translocation process. In order to understand this effect better a systematic study of the oscillating translocation time scaling behaviour with polymer length is required.

Recently there have been studies of polymer translocation in the presence of an oscillating driving force, $F(t)$ \cite{Nissila2012,Fiasconaro2010}. These are fundamentally different from the current study, as in the periodic forcing case the geometry of the pore is held fixed and therefore the effective translocation mobility will not change due to frictional effects with the environment. The transformation applied in Section IV is not applicable to periodic forcing and so a general form for the first passage time distribution may not be attainable.

A related case is where the mobility is spatially dependent and time independent, $\mu({\bf r})$. The effect of a spatially dependent mobility must be considered carefully as an extra drift terms arises in the equation of motion\cite{Schnitzer1993}. Finding a transform, as achieved in \Sec{diffusivegain}, to a mobility independent equation can become as hard as solving the original problem due the spatial derivatives and the difference in order between them.  Another active area of research is the pattern formation that arises from a mobility that is dependent on a local density field $\mu[\rho({\bf r}, t)]$\cite{Cates2010}.

In \Sec{polymer} we obtained mobilities through numerical simulations of polymer translocation. However, it is possible to consider the potential landscape of the pore to generate effective diffusion and translocations velocities\cite{Lubensky1999}. The effect of sequence heterogeneity could be included into the translocation dynamics by constructing a potential landscape, $U$, dependent on the translocation coordinate, $s$. Minimums in the landscape would correspond to complimentarily between the polymer structure and pore composition. The drift velocity would become modified to $\mu(-\nabla U(s) + {\bf F})$.

This work demonstrates some of the rich behaviour associated with time dependent changes of an environment. Measurements of the translocation time are typical in many experiments and, given a well defined device, the distribution of times could provide much information about a target. A component capable of frequency based selectivity would be a useful tool for integrated fluidic devices. An interesting future direction is to incorporate specific interactions along a channel to see if frequency based selectivity can be enhanced.

\appendix
\section{}
\label{sec:app}
The unit staircase function $\mathcal{S}(x)$ introduced in \Sec{twostate} has the following properties
\begin{eqnarray}
\mathcal{S}(x) &=& \lfloor x \rfloor  \\
		       &=& \sum_{n=1}^{\infty} \Theta(x-n) \\
\int_0^{y} \mathcal{S}(x)dx &=& \frac{\mathcal{S}^2(y-1)}{2} + \frac{\mathcal{S}(y-1)}{2} + y\mathcal{S}(y)\\
&-& \mathcal{S}^2(y) \nonumber\\ 
\frac{d\mathcal{S}(x)}{dx} &=& \sum_{n=1}^{\infty} \delta(x-n)\\
\lim_{a\rightarrow\infty} \mathcal{S}(ax) &=& ax
\end{eqnarray}

where $\lfloor x \rfloor$ is the floor operator, $\Theta(x)$ is the unit step function and $\delta(x)$ is the Dirac delta function.

%\appendix
\section{}
\label{sec:appb}

To construct the gain for non-diffusive periodic confinement we first split the total translocation time into two parts :
\begin{eqnarray}
t_{\rm osc} &=& t^\prime_{\rm osc} + \Delta t_{\rm osc}
\end{eqnarray}
where $t_{\rm osc^\prime}$ is the translocation time of the particle for $N_c$ complete cycles and $\Delta t_{\rm osc}$ is the translocation time in the remaining partial cycle required to complete the translocation event for length $L$. Therefore, $t^\prime_{\rm osc} = N_c T$ where $N_c = \mathcal{S}\left(L/\Delta X\right)$ and $\Delta X = F\int_0^T \mu(t) dt$ is the distance traveled per cycle. The average velocity per cycle is $\langle V \rangle_T = \Delta X/T$ and so $\langle t \rangle_T = L/\langle V\rangle_T  = LT/\Delta X$ and $\tilde{\omega} = \langle t \rangle_T/T = L/\Delta X$. Assimilating these relationships we have
\begin{eqnarray}
t^\prime_{\rm osc} &=& \frac{\mathcal{S}\left(\tilde{\omega}\right)}{\tilde{\omega}} \langle t \rangle_T
\end{eqnarray}
After traversing over $N_c$ complete cycles of time periodic confinement,  the remaining distance required to translocate will be $\Delta L = L - N_c \Delta X$. In dimensionless units we have
\begin{eqnarray}
\Delta\tilde{L} = \frac{\Delta L}{L} = 1 - \frac{\mathcal{S}(\tilde{\omega})}{\tilde{\omega}}
\end{eqnarray}
The distance travelled in time $t<T$ is $X(t) = F\int_0^t\mu(t)dt$ and the partial translocation time is obtained when the condition $X(\Delta t_{\rm osc}) = \Delta L$ is satisfied. Therefore, we need to solve the inverse problem to  obtain $\Delta t_{\rm osc}(X = \Delta L)$. In dimensionless units, $\Delta t_{\rm osc} = \Delta t_{\rm osc}(\Delta {\tilde L}) = \Delta {\tilde t}_{\rm osc}(\Delta {\tilde L}) \langle t \rangle_T$. The general formula for gain can now be expressed as in Eq. \ref{eq:ndpc}.

%\bibliography{aipsamp}
\end{document}